\documentclass[prb,aps,superscriptaddress,showpacs,floatfix,12pt]{revtex4}
\usepackage{amsmath}
\usepackage{amssymb}
\usepackage{color}
\usepackage{graphics}
\usepackage{epsfig}
\usepackage{hyperref}
\usepackage{epsfig}

\def\arccosh{{\rm arccosh}}

\begin{document}

\title{Surface Plasma Waves Across the Layers of Intrinsic Josephson Junctions}

\author{V.~A.~Yampol'skii}
\affiliation{Advanced Science Institute, The Institute of Physical
and Chemical Research (RIKEN), Wako-shi, Saitama, 351-0198, Japan}
\affiliation{A. Ya. Usikov Institute for Radiophysics and
Electronics National Academy of Sciences of Ukraine, 61085
Kharkov, Ukraine}

\author{D. R. Gulevich}
\affiliation{Advanced Science Institute,
The Institute of Physical and Chemical Research (RIKEN), Wako-shi,
Saitama, 351-0198, Japan}
\affiliation{Department of Physics, Loughborough University,
Loughborough LE11 3TU, UK} 

\author{Sergey Savel'ev}
\affiliation{Advanced Science Institute, The Institute of Physical
and Chemical Research (RIKEN), Wako-shi, Saitama, 351-0198, Japan}
\affiliation{Department of Physics, Loughborough University,
Loughborough LE11 3TU, UK}

\author{Franco Nori}
\affiliation{Advanced Science Institute, The Institute of Physical and
Chemical Research (RIKEN), Wako-shi, Saitama, 351-0198, Japan}
\affiliation{Department of Physics and MCTP, University of Michigan,
Ann Arbor, MI 48109-1120, USA}

\date{\today}
\begin{abstract}
{We predict surface electromagnetic waves propagating across the layers of intrinsic Josephson junctions. 
We find the spectrum of the surface waves and study the distribution of the electromagnetic field inside and outside the  superconductor. The profile of the amplitude oscillations of the electric field component of such waves is peculiar: initially, it increases toward the center of the superconductor and, after reaching a crossover point, decreases exponentially. 
}
\end{abstract}  \maketitle

\section{Introduction}
%
A very recent burst of interest to layered high-$T_c$ is due to the discovery of a new generation of  superconductors based on $\rm Fe As$ layers, ${\rm LaO_{1-x}F_{x}FeAs}$~\cite{LaO} and other such systems which possess a similar structure.
%
%
The conventional representative of layered high-$T_c$, ${\rm Bi_2 Sr_2 Ca Cu_2 O_{8+x}}$ (Bi2212) and ${\rm Tl_2 Ba_2 Ca Cu_2 O_{8+x}}$ (Tl2212) have a structure of superconducting $\rm Cu O_2$ layers with Josephson coupling between them.
The layered structures of high-$T_c$ favor the propagation of so-called Josephson plasma waves~\cite{JPW-plasma,matsuda}, propagating with frequencies above the Josephson plasma frequency $\omega_p$. 
The gap structure of the Josephson plasma excitation spectra has been experimentally observed from measurements of the Josephson plasma resonance~\cite{matsuda}. 
Josephson plasma waves can exhibit remarkable features, including the slowing
down of light, self-focusing effects~\cite{natphys}
and are linked to applications in the THz frequency range~\cite{applic}.

It was recently predicted that the layered structure of high-$T_c$ superconductors allows the propagation of surface waves~\cite{Saveliev-surface}. 
Such waves propagate below the Josephson plasma frequency $\omega_p$ and propagate in the vicinity of the superconducting surface {\it along} the layers. 
In this paper we show that there exist surface electromagnetic TM-waves propagating {\it across} the superconducting layers.

The electric, $\mathbf{E} = \{E_x, 0,E_z\}$, and magnetic, $\mathbf{H} = \{0, H, 0\}$, components of the electromagnetic waves are proportional to $\exp[i\,(q\,x - \omega\, t)]$ and decay both in the vacuum and inside the layered superconductor. 
Such surface waves across the layers are strongly influenced by an external magnetic field $\mathbf{h}_0$ applied along the superconducting layers. Here we describe the propagation across the layers of such surface waves and estimate the influence of an external magnetic field on their spectrum.

\begin{figure}[hbpt]
\includegraphics[width=3.5in]{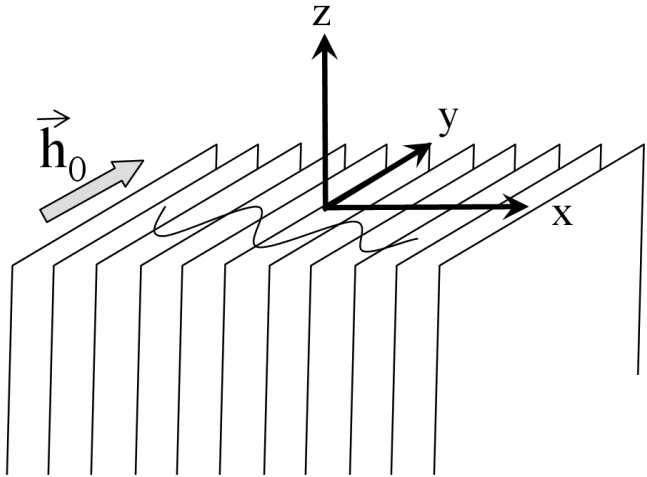}
\caption{\label{geometry} Interface between vacuum ($z>0$) and a layered superconductor ($z<0$) in an external magnetic field $\mathbf{h}_0$. A layered high-$T_c$ superconductor has a structure of superconducting layers coupled via intrinsic Josephson junctions.
}
\end{figure}

\section{Model and Results}
Consider an interface between the vacuum ($z>0$ in Fig.~\ref{geometry}) and a layered superconductor ($z\le 0$). Let the c-axis of the superconductor be along the $x$-axis 
so that the vacuum-superconductor interface lies in the $xy$-plane and
an external magnetic field $\mathbf{h}_0$ is applied along the $y$-axis, parallel to the superconducting layers, see Fig.~\ref{geometry}. 
The electromagnetic field inside the layered superconductor ($z<0$) is determined by the distribution of the gauge invariant phase difference $\varphi(x,z,t)$ of the order parameter between neighboring layers. 
It is described by a set of coupled sine-Gordon equations~\cite{SG-coupled},
that in the continuum limit (see, e.g.,~\cite{Saveliev-generation}) can be written as,
\begin{equation}
\label{s-G}
\left(1-\lambda_{ab}^2\frac{\partial^2}{\partial
x^2}\right)\left(\frac{\partial^2
    \varphi}{\partial t^2}+\omega_p^2 \sin\varphi\right)
    -\lambda_{c}^2 \omega_p^2 \frac{\partial^2 \varphi}{\partial z^2}=0 .
\end{equation}
Here we neglect the relaxation terms caused by the quasiparticle
conductivity; $\lambda_{ab}$ and
$\lambda_{c}=c/\omega_p\sqrt{\varepsilon}$ are the magnetic penetration depths across and along layers, respectively,
$\omega_p = (8\pi e D j_c/\hbar\varepsilon)^{1/2}$ is the
Josephson plasma frequency. The latter is determined by the
critical Josephson current $j_c$,  the interlayer dielectric
constant $\varepsilon$, and the spatial period of the layered structure $D$.
The gradient of the superconducting phase is related to the magnetic field $h(z)$, directed along $y$, as (e.g.,~\cite{Koshelev-Aranson})
\begin{equation}
-\frac{\partial\varphi}{\partial z}=\frac{2\pi D}{\Phi_0}\left(1-\lambda_{ab}^2\frac{\partial^2}{\partial x^2}\right) h(z)
\label{phi-gradient}
\end{equation}
Using~\eqref{phi-gradient} as a boundary condition at $z=0$, we solve Eq.~(\ref{s-G}) to obtain the dependence of the superconducting phase $\varphi (z)$ on the distance $z$ from the interface
under a homogeneous stationary magnetic field $h_0$,
\begin{equation}
\label{phi0}
\varphi_0(z)=-4\arctan\left[\exp\left(\frac{z-z_0}{\lambda_c}\right)\right],
\quad z<0,
\end{equation}
where a positive constant $z_0>0$ is defined by the boundary condition
\begin{equation*}
-\,\frac{\partial\varphi_0(z)}{\partial z}|_{z=0}=\frac{2\pi D}{\Phi_0}\,h_0
\end{equation*}
so that
\begin{equation*}
z_0=\lambda_c\,\arccosh\left(\frac{h_c}{h_0}\right),
\quad\text{where}\quad 
h_c=\frac{\Phi_0}{\pi D\lambda_c}.
\end{equation*}
Here we study the case of relatively small fields, when $h_0$ is
less than the critical value $h_c$ and Josephson vortices do not penetrate the superconductor. 

\subsection{Surface waves at $h_0\lesssim h_c$}
We take into account the $t$ and $x$ dependence of superconducting phase $\varphi(x,z,t)$ as small variations around the stationary configuration $\varphi_0(z)$ given by Eq.~\eqref{phi0}. Assuming 
$$\varphi (x,z,t)=\varphi_0(z)+\varphi_w(x,z,t),$$
as a sum of the static and wave terms, we linearize the Eq.~(\ref{s-G}) 
to obtain
\begin{equation*}
\left(1-\lambda_{ab}^2\frac{\partial^2}{\partial
x^2}\right)\left(\frac{\partial^2
    \varphi_w}{\partial t^2}+\omega_p^2\,\varphi_w\, \cos\varphi_0(z)\right)
    -\lambda_{c}^2 \omega_p^2 \frac{\partial^2 \varphi_w}{\partial z^2}=0 .
\end{equation*}
Substituting $\varphi_w(x,z,t)=\xi(z)\exp(i\,q\,x - i\,\omega\, t)$ we derive an ordinary differential equation for~$\xi(z)$,
\begin{equation}
\label{eq}
  -\frac{\lambda_c^2}{(1+q^2\lambda_{ab}^2)}\,\frac{d^2\xi}{dz^2}+
  \left(1-\Omega^2-\frac{2}{\cosh^2[(z-z_0)/\lambda_c]}
  \right)\xi=0,
\end{equation}
where we have introduced $\Omega\equiv\omega/\omega_p$. Here we are interested in a solution decaying inside the layered superconductor: $\xi(z)\rightarrow 0$ at $z\rightarrow-\infty$.
The equation~(\ref{eq}) has the form of a 1D Schr\"{o}dinger equation for a particle 
with energy 
$$E(\Omega)=\Omega ^2-1$$
in a potential 
$$U(z)=-\,\frac{2}{\cosh^2[(z-z_0)/\lambda_c]}.$$ 
The bound states corresponding to the waves decaying at $z\rightarrow -\infty$, can exist for negative energies $E(\Omega)<0$, i.e. for $\Omega<1$.
One can write an exact solution of Eq.~(\ref{eq}) in terms of the Hypergeometric function, 
$$
\xi(z)=\left(1-\zeta(z)^2\right)^{\epsilon/2}\,F\left(\epsilon-s,\, \epsilon+s+1,\, \epsilon+1,\, \frac{1+\zeta(z)}{2}\right)
$$
where
$$
\zeta(z)=\tanh\left(\frac{z-z_0}{\lambda_c}\right)
$$
and
$$
s=\frac12\,\left(-1+\sqrt{1+8(1+q^2\lambda_{ab}^2)}\right)
$$
$$
\epsilon=\sqrt{(1-\Omega^2)(1+q^2\lambda_{ab}^2)}
$$
We have studied the behavior of the spectrum of surface Josephson plasma waves
by means of the WKB approximation valid for 
$$Q\equiv q\lambda_{ab}\gg 1.$$ 
If the inequalities
\begin{equation*}
0\,<\,(1-\Omega^2)\,<\frac{2h_0^2}{h_c^2}
\end{equation*}
are satisfied, there exists a classical turning point $z=z_t$.
According to Eq.~(\ref{eq}), this point is defined by the equation $E(\Omega)=U(z_t)$ that leads to
\begin{equation*}
\label{tp}
  1-\Omega^2=\frac{2}{\cosh^2[(z_t-z_0)/\lambda_c]}.
\end{equation*}
The ``wavefunction'' $\xi(z)$ oscillates in the region
$z_t<z<0$ and exponentially decays at $-\infty <z<z_t$. 
After the procedure of matching the ``wavefunctions'' at the turning point by the connecting formulas known from quantum mechanics, we obtain the quasiclassical expression for $\xi(z)$. For the classically-allowed region $z_t<z<0$, we have
\begin{equation}
 \xi(z) \simeq
 \frac{A}{\left[E(\Omega)-U(z)\right]^{1/4}}\,
 \cos\left[\frac{\sqrt{1+Q^2}}{\lambda_c}
 \int_{z_t}^zdz'
\sqrt{E(\Omega)-U(z')}-\frac{\pi}{4}\right].
\label{psi-w}
\end{equation}
and the underbarrier ``wavefunction'' for $-\infty<z<zt$,
\begin{equation*}
 \xi(z) \simeq
 \frac{A/2}{\left[U(z)-E(\Omega)\right]^{1/4}}\,
 \exp\left[\frac{\sqrt{1+Q^2}}{\lambda_c}
 \int_{z_t}^zdz'
\sqrt{U(z)-E(\Omega)}\right].
\end{equation*}
The waves $\xi(z)\exp(i\,q\,x - i\,\omega\, t)$, corresponding to the solution~\eqref{psi-w}, are Josephson plasma waves running along the interface of the layered superconductor and {\it across} its layers. From Eq.~\eqref{phi-gradient} we obtain the relation of $\xi(z)$ to amplitudes of the electromagnetic field components in the layered superconductor. For the magnetic field, distribution inside the sample, we obtain:
\begin{equation}
H(z)= - \,\frac{h_c\lambda_c}{2(1+q^2\lambda_{ab}^2)}\frac{d\xi}{dz}.
\label{H}
\end{equation}
From the ac Josephson relation, Maxwell equations, and substituting $\lambda_c=c/\omega_p\sqrt{\epsilon}$ and $h_c=\Phi_0/\pi D\lambda_c$, we obtain the amplitudes of the electric field components:
$$
E_x(z)=\frac{\Phi_0}{2\pi\,c\,D}(-i\omega)\xi(z)=-i\frac{h_c\Omega}{2\sqrt{\varepsilon}}\,\xi(z),
$$
and
$$
E_z(z)=\frac{\lambda_{ab}^2\, q\, \Omega}{\lambda_c\sqrt{\varepsilon}}\,H(z)
$$
Using the Maxwell equations in vacuum, we obtain the dependence of the electromagnetic field components outside the superconductor. This gives an exponential decay for positive $z$,
\begin{equation*}
H^{\rm vac}, \, E_x^{\rm vac},\, E_z^{\rm vac} \propto
\exp(iqx-i\omega t-k_v z), \quad z >0
\end{equation*}
with the decay constant $k_v = \sqrt{q^2-\omega^2/c^2}>0$ for $q>\omega/c$.
In the WKB regime, when $Q\gg1$, we obtain 
$$
k_v = \frac{1}{\lambda_{ab}}\sqrt{Q^2-\frac{\lambda_{ab}^2\Omega^2}{\lambda_c^2\varepsilon}}\simeq\frac{Q}{\lambda_{ab}}
$$
as $\lambda_{ab}/\lambda_c\varepsilon \ll 1$. Because of the large $\lambda_c/\lambda_{ab}\gg1$ and $Q\gg1$, the surface wave decays very quickly in vacuum, on the scale $\sim\lambda_{ab}/Q$, which is much smaller than $\lambda_c$. 

\begin{figure}[h]
\includegraphics[width=4.5in,clip]{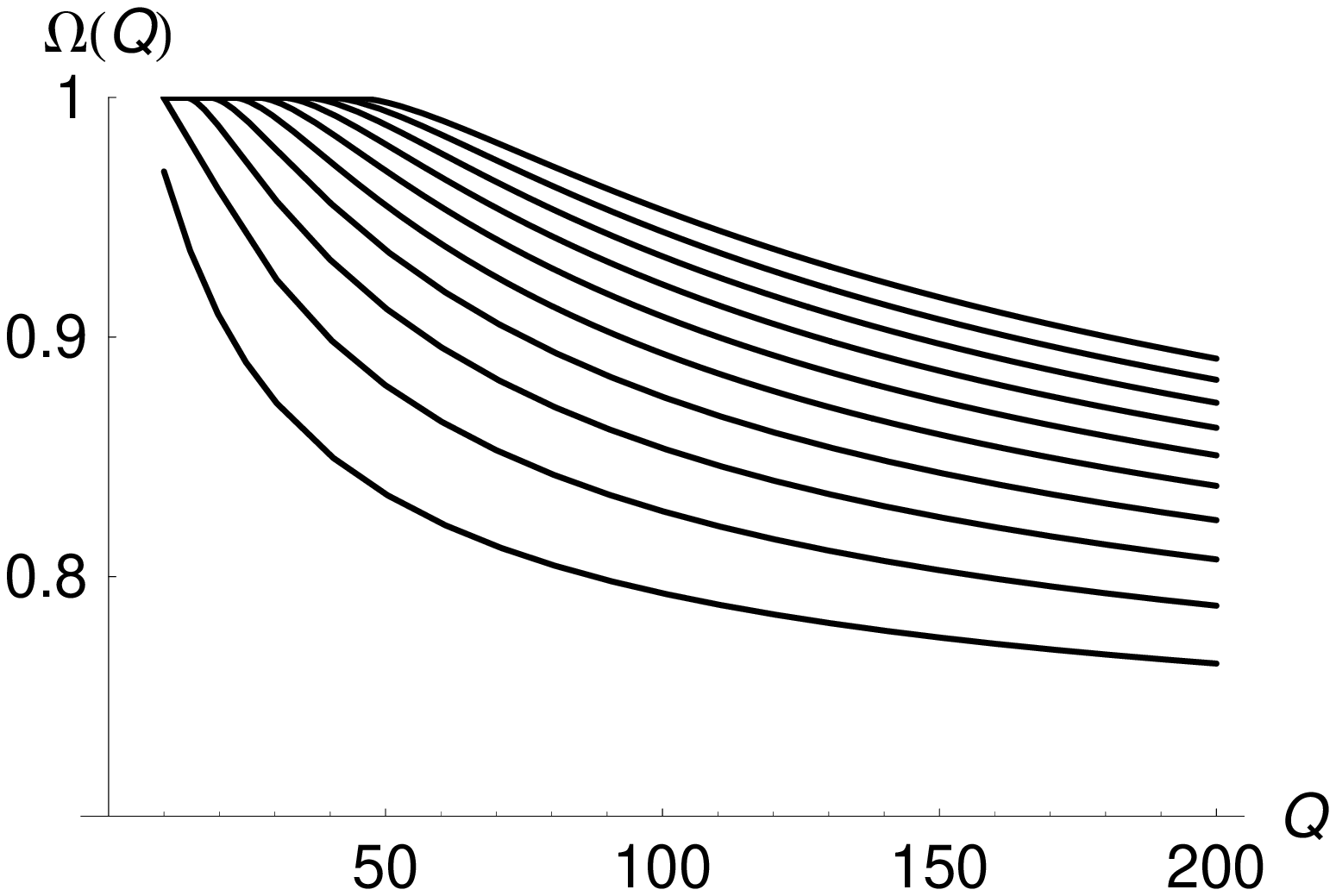}
\includegraphics[width=4.5in,clip]{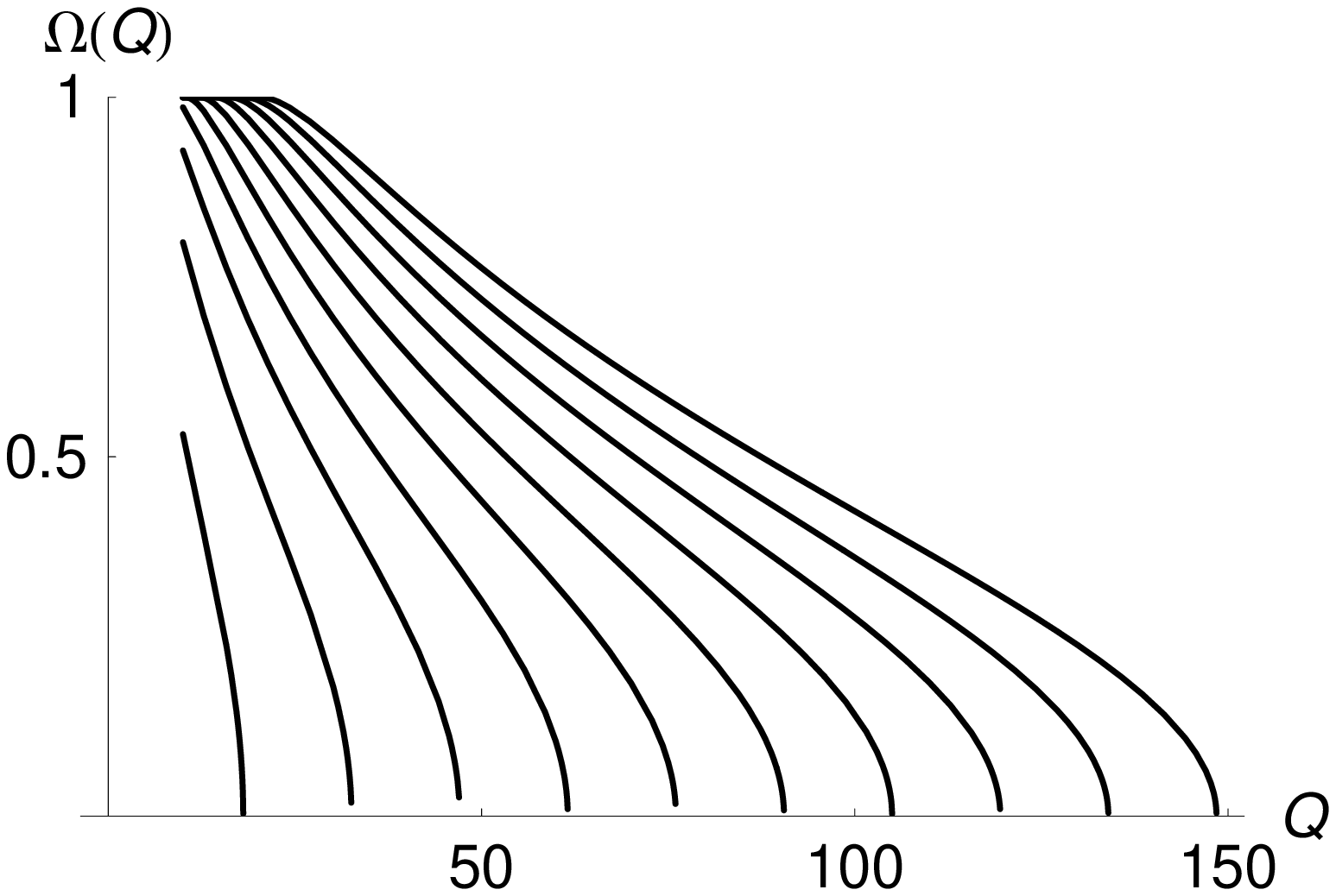}
\caption{\label{dispersion} (a) Dispersion curves for surface Josephson plasma waves at an external magnetic field $h_0/h_c=0.5$. The branches correspond to $n=1,2,...,10$ from bottom to top. The spectrum is limited from below by the value $\Omega>(1-2h_0^2/h_c^2)^{1/2}\simeq 0.71$  (b) Dispersion curves for surface Josephson plasma waves at an external magnetic field $h_0/h_c=0.9$. The branches correspond to $n=1,2,...,10$, from left to right.} 
\end{figure}

The ratio of amplitudes for the tangential electric and magnetic fields at the interface $z=+0$,
above the surface of superconductor, is
\begin{equation}\label{v-imp}
\frac{E_x^{\rm vac}}{H^{\rm vac}}= \frac{i c}{\omega}k_v =\frac{i
c}{\omega}\sqrt{q^2-\omega^2/c^2}.
\end{equation}
In order to derive the dispersion relation for surface Josephson plasma waves, we calculate
the ratio $E_x(0)/H(0)$ in the superconductor using Eqs.~(\ref{H}) and (\ref{psi-w}) and
then equate this ratio to the vacuum impedance Eq.~(\ref{v-imp}). This gives
\begin{equation*}
\frac{\lambda_{ab}\Omega^2\, \sqrt{1+Q^2}}{\lambda_c\varepsilon\,
\sqrt{\left(Q^2-\frac{\lambda_{ab}^2\Omega^2}{\lambda_c^2\varepsilon}\right)\left(\frac{2h_0^2}{h_c^2} + E(\Omega)\right)}} =  \tan\left[-\,\frac{\sqrt{1+Q^2}}{\lambda_c}
\int_{z_t}^0 dz'
\sqrt{E(\Omega)-U(z')}+\frac{\pi}{4}\right]
\end{equation*}
with $Q=q\lambda_{ab}$. Because 
$\lambda_{ab}/\lambda_c\varepsilon \ll 1$, 
this relation can be simplified disregarding the vacuum contribution. Thus,
\begin{equation}
\frac{\sqrt{1+Q^2}}{\lambda_c}
\int_{z_t}^0 dz'
\sqrt{E(\Omega)-U(z')} =\pi\left(n+\frac14\right), \quad n=1,2,3 \dots.
\label{as-dr}
\end{equation}
A set of dispersion curves for $n=1,...,10$ is shown in Fig.~\ref{dispersion} for two different values of the external magnetic field, $h_0/h_c=0.5$ (Fig.~\ref{dispersion}a) and $h_0/h_c=0.9$ (Fig.~\ref{dispersion}b).

\begin{figure}[hbpt]
\includegraphics[width=4.5in]{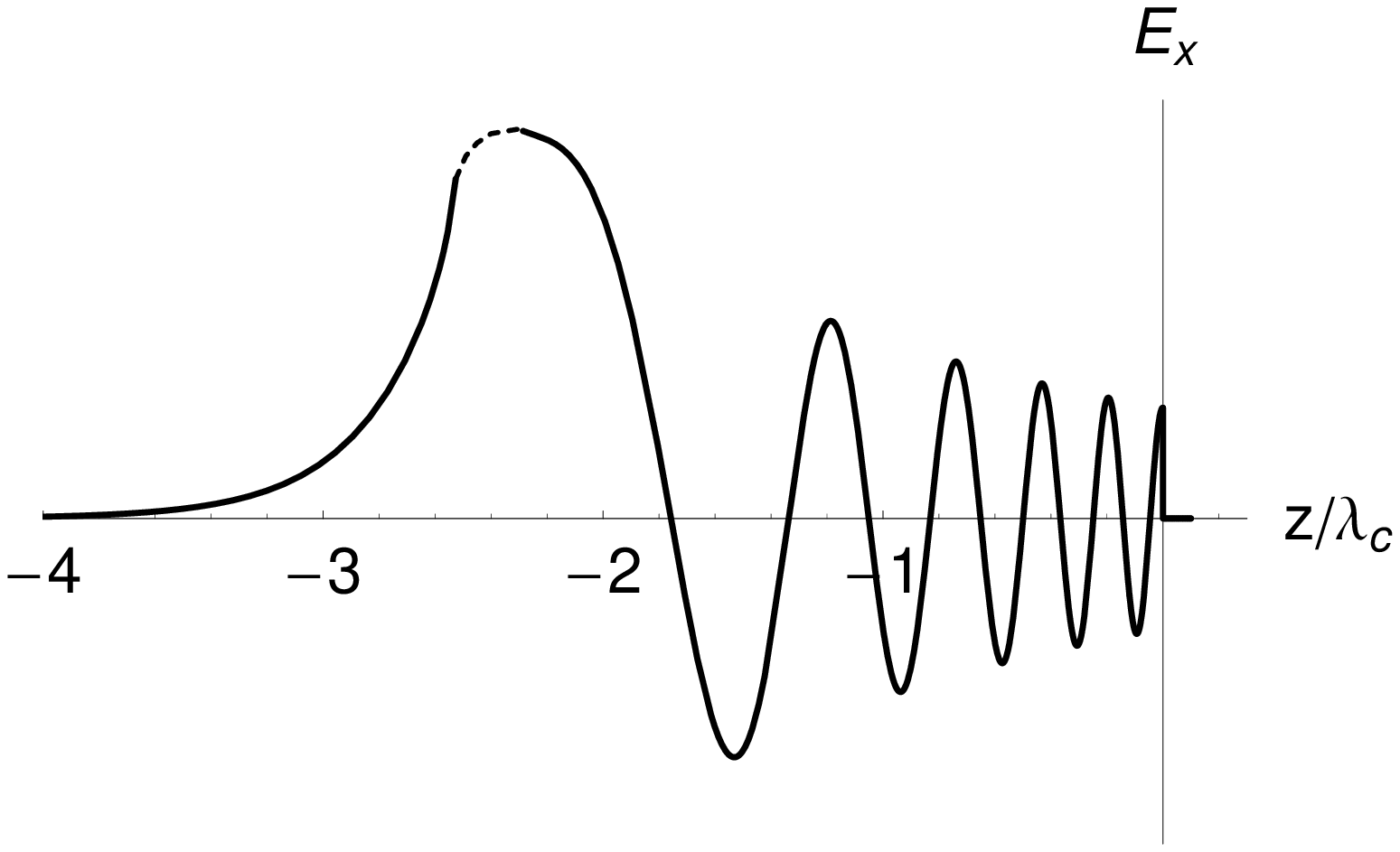}
\caption{\label{f3} Spatial distribution of the field $E_x(z)$ of surface Josephson plasma waves calculated for $h_0/h_c=0.5$, $n=10$, $Q=50$. The dashed curve corresponds to the classical turning point $z=z_t$, where the WKB approximation fails. $z<0$ corresponds to a plasma wave in the layered superconductor, while $z>0$ corresponds to vacuum.
As seen from the figure, the amplitude of the surface wave decays very quickly in vacuum on a scale much smaller than $\lambda_c$.
Notice that the number of nodes of the distribution $E_x(z)$ corresponds to the integer parameter $n$ in Eq.~\eqref{as-dr}.}
\end{figure}

The dispersion relation Eq.~(\ref{as-dr}) corresponds to surface waves of an unusual nature. The electromagnetic
field does not decrease monotonically into the superconductor. Instead, the number of
oscillations of $\xi(z)$ with \emph{increasing amplitude} occur before the exponential decrease. An example of the
oscillating field $E_x(z)$ distribution in surface Josephson plasma waves with
for parameters $h_0/h_c=0.5$, $n=10$ and $Q=50$ is shown in Fig.~\ref{f3}.

\begin{figure}[hbpt]
\includegraphics[width=4.5in]{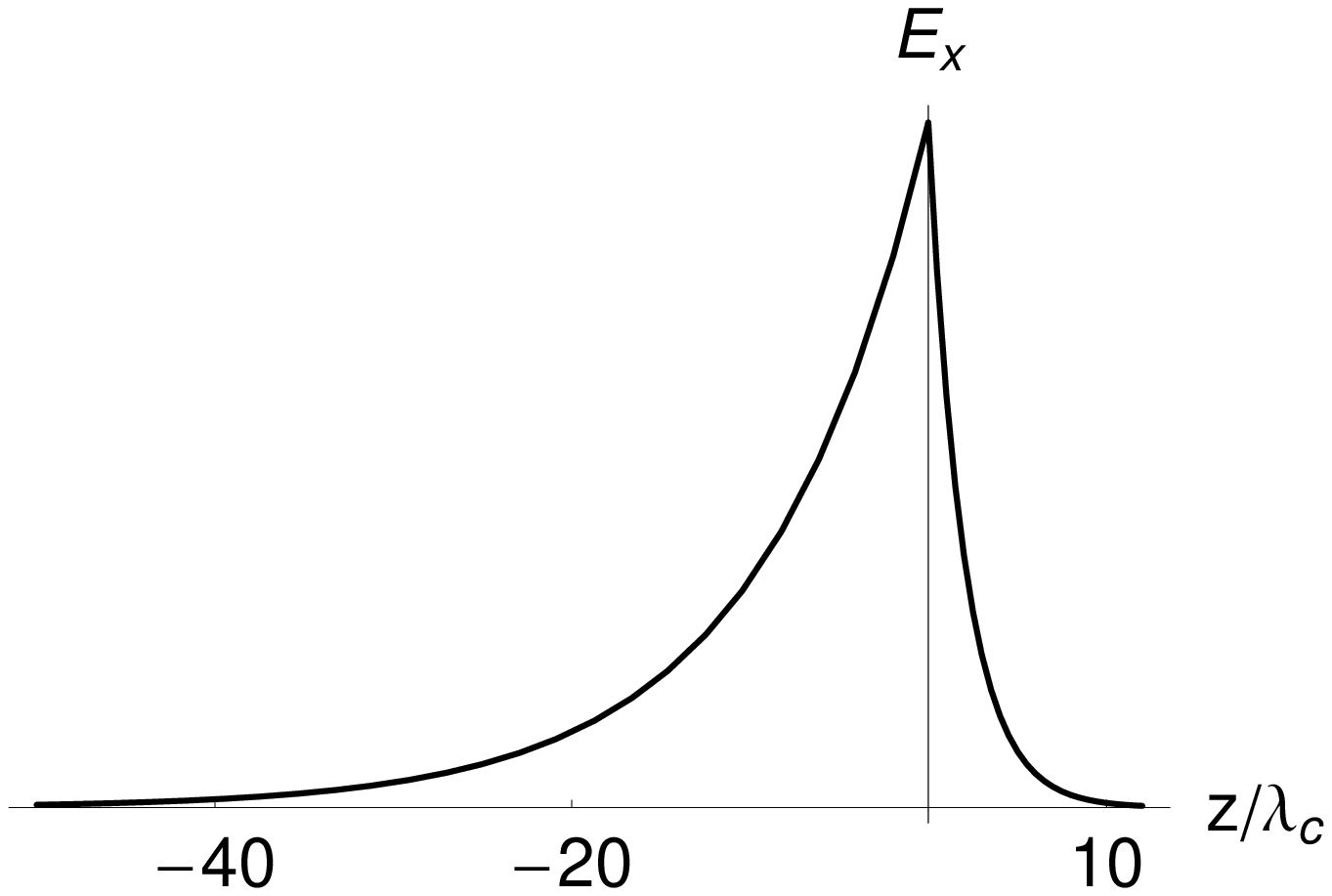}
\caption{\label{f5} Spatial distribution of the field $E_x(z)$ of surface Josephson plasma waves calculated for $h_0=0$ and $Q=0.001$. $z<0$ corresponds to a plasma wave in the layered superconductor, while $z>0$ corresponds to the vacuum.}
\end{figure}

\begin{figure}[hbpt]
\includegraphics[width=4.5in]{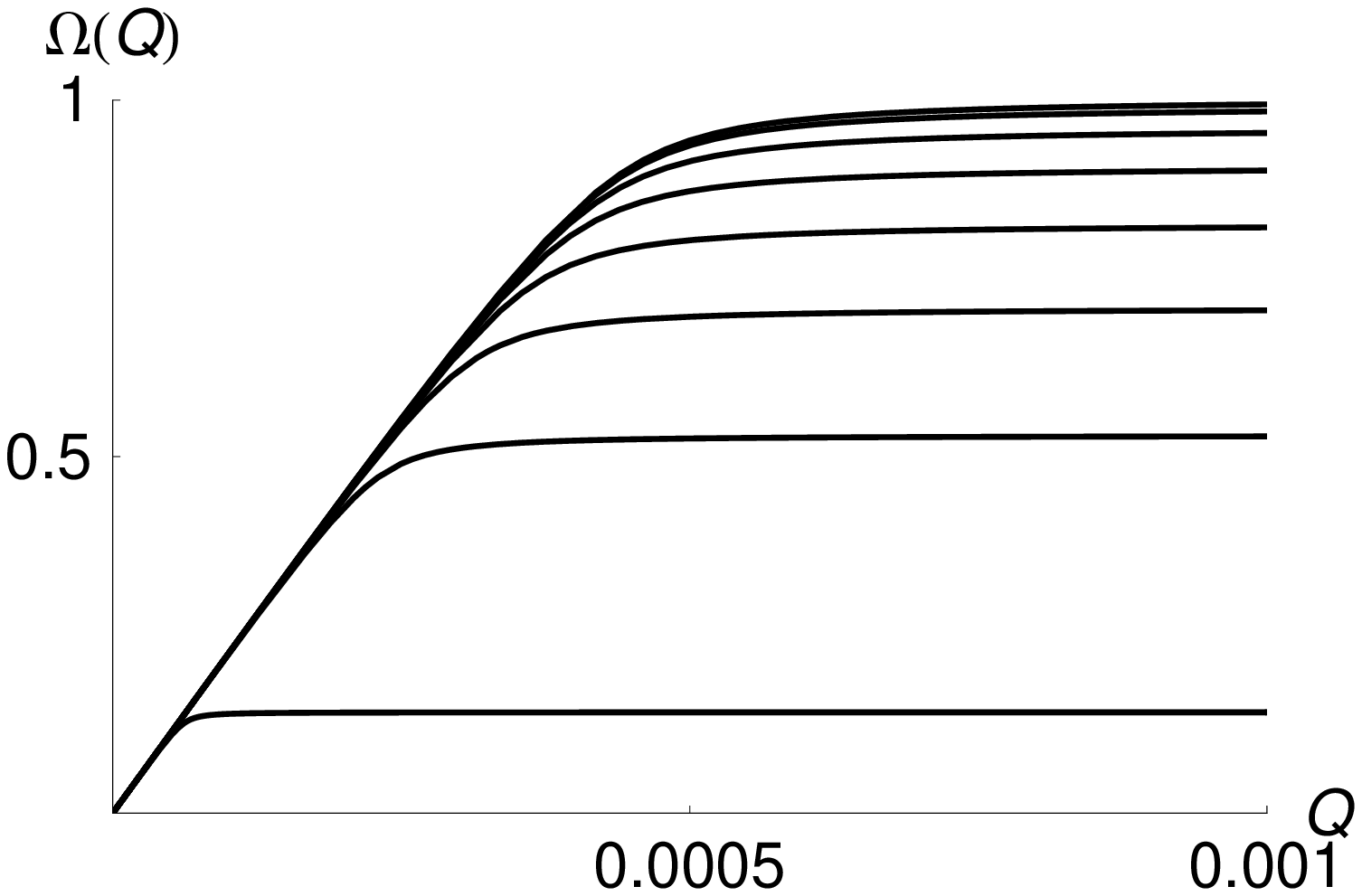}
\caption{\label{0-branch} Dispersion curve for surface Josephson plasma waves at different values of an external magnetic field given by Eq.~\eqref{dr2}. The branches correspond to magnetic field values $h_0/h_c=0,0.1,0.2,...,0.7$ numbered from top to bottom. } 
\end{figure}

\subsection{Surface waves at small $h_0$}
The equation~(\ref{as-dr}) does not describe all branches of the
spectrum of surface Josephson plasma waves. For relatively small magnetic fields and small frequency $\Omega$, when the inequality
\begin{equation*}
 E(\Omega)=\Omega^2-1<-\frac{2h_0^2}{h_c^2}
\end{equation*}
is fulfilled, the whole superconducting area $z<0$ is classically
disallowed. In this case, the quasiclassical wave function
$\xi(z)$ exponentially decreases starting from the boundary,
\begin{equation*}
 \xi(z) \simeq
 \frac{A}{\left[U(z)-E(\Omega)\right]^{1/4}}
 \exp\left[\frac{\sqrt{1+Q^2}}{\lambda_c}
 \int_0^zdz'
\sqrt{U(z')-E(\Omega)}\right].
\end{equation*}
A typical distribution of the wave amplitude for $Q=0.001$ and $h=0$ is shown in Fig.~\ref{f5}. The corresponding dispersion relation can be written as
\begin{equation}
\label{dr2}
\sqrt{Q^2-\frac{\lambda_{ab}^2\Omega^2}{\lambda_c^2\varepsilon}}=
\frac{\lambda_{ab}\Omega^2\sqrt{1+Q^2}}{\lambda_c\varepsilon
\sqrt{-E(\Omega)-2h_0^2/h_c^2}}.
\end{equation}
This dispersion curve is shown in Fig.~\ref{0-branch} for several values of the magnetic field, $h_0/h_c=0,0.1,...,0.7$. We have used the parameters: $\lambda_c/\lambda_{ab}=500$, and $\varepsilon=20$. Note
that Eq.~(\ref{dr2}) is valid not only in the WKB approximation
and applicable even in the absence of an external magnetic field.

\section{Conclusion}
We have predicted Josephson plasma waves propagating along the surface of anisotropic high-$T_c$ and {\it across} its  superconducting layers. There exist different modes of such waves ordered by the number of nodes $n$ of the amplitude of the electric field component inside the superconductor. The profile of the surface waves is unusual: first, the amplitude of the  oscillations increases inside the superconductor and, after reaching the last node, decreases exponentially.


\section{Acknowledgements}
We gratefully acknowledge partial support from the National
Security Agency (NSA), Laboratory Physical Science (LPS), Army Research Office (ARO), National Science
Foundation (NSF) Grant No.~EIA-0130383, Core-to-Core
(CTC) program supported by the Japan Society for the Promotion
of Science (JSPS), JSPS-Russian Foundation for Basic
Research (RFBR) No.~06-02-91200.
Also, we acknowledge support from the Ministry of Science,
Culture and Sports of Japan via the Grant-in Aid for Young
Scientists No. 18740224, the EPSRC via Nos. EP/D072581/1 and EP/F005482/1, 
Postdoctoral Research Fellowship EP/E042589/1 and ESF network programme
"Arrays of Quantum Dots and Josephson Junctions."


\end{document}